%% file: main.tex
\documentclass[sigconf,nonacm]{acmart}
\renewcommand\footnotetextcopyrightpermission[1]{}
\newcommand{\ind}{\mathbf{1}}
\makeatletter
\let\tableinput\@@input
\makeatother
\newenvironment{appendixtable}{\par\vskip\medskipamount\noindent\begin{minipage}{\textwidth}\captionsetup{type=table}}{\end{minipage}\par\vskip\medskipamount}
\begin{document}
\title[The Company a Question Keeps]{The Company a Question Keeps:\texorpdfstring{\\}{ }Auditing Open Jev-Style Models for Retrieval}
\author{Jianbo Chen}
\email{cjb24@mails.tsinghua.edu.cn}
\affiliation{%
  \institution{Tsinghua University}
  \city{Beijing}
  \country{China}}
\begin{abstract}
Jev's documented interface evaluates typed questions independently against shared evidence. We audit request independence in seven open Jev-style checkpoints on 276 constructed three-passage states from 92 TREC retrieval queries. Paired interventions separately change requested questions, displayed evidence, and candidate eligibility. Request expansion induces strict binary flips in 9.1\%/8.0\% of Nimble's judgments and changes 21.4\%/18.9\% of selection-or-abstention actions in TREC DL2019/2020 under a threshold of 0.5 fixed before inference. Correct-action rates change by 0.0/+8.2 percentage points, respectively, with 95\% query-bootstrap intervals of $[-6.0,6.0]$ and $[3.8,12.6]$. Field masking traces all 70 request-set flips to the schema fields serialized into each scoring prompt. In Von, fixing a shared separator's position eliminates two eligibility-induced preference flips in selected diagnostic states. Laya-typed's default CUDA path produces five passage swaps without strict binary flips, and both yearly action-change intervals include zero; in an FP32 rerun, its probability movement vanishes at the model's four-decimal reporting resolution.
\end{abstract}
\maketitle
\raggedbottom
\let\balance\relax

\section{Introduction}
Retrieval systems use model outputs to judge whether passages answer a query, select a candidate, or abstain. These decisions can take different forms: categorical selection distributes probability over competing options, binary judgments can accept several passages, and an ordinal rubric assigns relevance grades.

Jev separates a shared \emph{state} from typed \emph{questions}. Its documented interface evaluates questions independently against that state \citep{typesafeintro}. We use request independence as an audit target for open Jev-style implementations. An autoregressive implementation can expose each judgment to other requested questions in the prompt and to earlier generated answers. Output-format validation leaves these dependencies unchecked.

Our audit separates three changes that a retrieval application can make: ask another question, add a passage to the evidence, or make a passage eligible for selection. Asking a question about an already displayed passage B leaves the evidence for passage A unchanged. Adding passage B to the input changes the evidence; making a displayed passage selectable changes the eligible set.

Our exploratory audit evaluates seven open, trained checkpoints through four released inference engines: two Laya variants, three Kev scales, Bespoke Nimble, and Von. Four ordinary-LLM checkpoints serve as controls through three computations: a question-isolated compiler, a schema-visible single-answer probe, and constrained joint generation. All use the same 92 retrieval queries and three answer-count scenarios.

The audit links request dependence in Nimble to serialized schema fields and eligibility dependence in Von to a shared separator's position on selected states. A precision diagnostic examines Laya's small batch-dependent probability shifts. Fixed selection and abstention rules relate these probability changes to binary judgments and task correctness.

\section{Related Work}
\paragraph{Decision-model evaluation and request conditions.}
Jev's interface provides categorical selection (Choice), yes/no probability (Noul), and ordinal assessment (Score) \citep{typesafechoice,typesafenoul,typesafescore}. Zhang et al.\ hold legal documents and target judgments fixed while changing hypothesis visibility, requested outputs, and output order; they also separate corrections, regressions, and stable errors \citep{zhang2026samescores}. Guo et al.\ evaluate Jev as an alignment-failure detector \citep{guo2026justaskjev}. JEVal compares decision models and generative LLMs across domains and examines their behavior in larger systems \citep{duan2026jeval}. Our retrieval audit locates the input channels responsible for observed dependencies through interventions on released implementations.

\paragraph{Contextual retrieval decisions.}
Pointwise rerankers score a query--document pair \citep{nogueira2019passage}, whereas listwise prompting exposes candidate lists \citep{sun2023chatgptsearch} and pairwise prompting decomposes ranking into local comparisons \citep{qin2024pairwise}. Huang et al.\ study how candidate composition and order affect relevance judgments \citep{huang2026contextual}. JevOut optimizes natural context additions to redirect initially correct decisions \citep{xu2026jevout}. Our evidence additions use preselected passages, and the request-set comparisons hold the full passage state fixed.

\paragraph{Candidate interfaces and rejection.}
Option labels and positions affect multiple-choice selection and preferential rankings \citep{zheng2024mcqselectors,zhao2024inconsistency}. Sun et al.\ test option-name/definition binding in typed decision models \citep{sun2026typesafe}; Joy compares direct and hierarchically reconstructed probabilities in Jev and Laya \citep{joy2026coherence}. Zhong et al.\ study Jev's rejection of answer-absent arithmetic menus \citep{zhong2026missing}, and Lu and Wu evaluate candidate coverage and rejection-policy transfer \citep{lu2026coverage}. Our eligibility contrast adds C to the selectable set while holding displayed passages and retained A/B criteria fixed. We use conditional A/B preference to test Von's positional dependency.

\paragraph{Numerical batch dependence.}
Batch-invariant kernels support reproducible LLM inference \citep{he2025nondeterminism}. Our Laya diagnostic checks whether the observed request contrast persists in FP32 and records its consequences for a fixed retrieval decision rule.

\section{Audit Definitions}
Let $s$ contain a retrieval query and displayed passages, $Q$ denote the ordered collection of requested typed questions, and $q_i\in Q$ denote the target question. Its instruction asks whether passage $i$ itself contains information answering the retrieval query.

\subsection{Computation and input visibility}
Laya, Kev, Nimble, and Von use their released decision weights and inference engines. To describe their dependencies and construct ordinary-LLM controls, we distinguish three ways of computing the requested answers.
\paragraph{Independent question compilation.}
The compiler produces $x_i=C(s,q_i)$, omitting other questions and the target's routing key. For fixed model parameters and deterministic inference, its output is
\begin{equation}
 p_i^{\mathrm{ind}}=F(C(s,q_i)).
\end{equation}
Holding $s,q_i$ fixed makes $x_i$ identical across request sets, giving request invariance by construction. Changing the state to $s'$ changes the input. A document-isolated control removes the other passages as well.

\paragraph{Schema-visible probe.}
The probe serializes the state, a question list, and a target key, then asks for only that target's label:
\begin{equation}
 p_i^{\mathrm{vis}}=F(C_{\mathrm{vis}}(s,Q,q_i)).
\end{equation}
Each target is evaluated separately, with the other questions and routing keys visible but their answers absent. Jev itself keeps routing keys outside the model input \citep{typesafechoice}.

\paragraph{Joint autoregressive output.}
The joint realization asks for one Yes/No answer per line. With $h_i=(s,Q,\hat y_{<i})$ and $L=\{\mathrm{Yes},\mathrm{No}\}$, it scores
\begin{equation}
 p_i^{\mathrm{joint}}=\frac{P_\theta(\mathrm{Yes}\mid h_i)}{\sum_{l\in L}P_\theta(l\mid h_i)}.
\end{equation}
This probability is normalized over the two allowed labels, rather than the full vocabulary. After each decision, the winning label and a forced newline are appended. Later scores therefore condition on earlier generated answers.

\subsection{Output support and probability}
Given allowed single-token labels $L$, logits $z_l$, and vocabulary $V$, we retain both
\begin{equation}
 \widetilde p_l=\frac{e^{z_l}}{\sum_{k\in L}e^{z_k}},\qquad
 M_L=\frac{\sum_{k\in L}e^{z_k}}{\sum_{v\in V}e^{z_v}}.
\end{equation}
$\widetilde p$ is the distribution normalized over allowed labels; $M_L$ measures the model's support for emitting that alphabet. Low $M_L$ flags a mismatch between the requested output format and the model's next-token distribution.

\subsection{Judgments and actions}
Noul-style judgment uses Yes/No labels. Binary Choice encodes the same proposition using true/false criteria and A/B labels. Score uses an ordinal 0--3 rubric; we compute its positive mass $\widetilde p_2+\widetilde p_3$. Selection Choice offers passage labels A/B/C, optionally adding N for none. The binary-Choice comparison changes label encoding and instructions; the ordinal comparison additionally changes the rubric.

For binary scores $p_i$, the fixed workflow returns the highest-scoring passage with $p_i>t$, breaking ties by label; it abstains if none passes. Thresholds $t\in\{0.5,0.7,0.9\}$ are fixed before inference. Selecting a grade-2/3 passage is a correct action; abstention is correct only if the state has no such passage. For the state-addition contrast, the program selects among the original three targets; the new passage changes only the evidence.

\subsection{Outcome measures}
For paired binary scores $p_i,p_i'$, we measure $|p_i'-p_i|$, the strict flip indicator $\ind[(p_i-0.5)(p_i'-0.5)<0]$, and the change in thresholded correctness. Pairs with either score exactly at 0.5 do not count as strict flips and are recorded separately. A program-level action change occurs when the selected passage or abstention status changes. This can occur without a strict flip if two accepted passages exchange order. For Choice eligibility, retained A/B preference is determined by the sign of the A/B logit difference or, equivalently, whether the conditional A/B probability exceeds 0.5. We measure probability movement using that conditional probability, since expanding the allowed alphabet mechanically changes the denominator of the raw A probability.

\section{Experimental Design}
\subsection{Data and chronology}
We intersect official TREC DL2019/2020 passage relevance judgments (qrels) with the supplied MS MARCO top-1000 text \citep{craswell2019trecdl,bajaj2016msmarco}. Exact text duplicates are removed within each query. Eligibility requires two unique grade-2/3 passages and four grade-0 passages, yielding 39 queries in 2019 and 53 in 2020. Grade 1 and unjudged passages are not used. Eligible passages are selected in source-file order without consulting model outputs.

Each query contributes three states of three passages each, with zero, one, or two grade-2/3 (positive) passages. Passage order and A/B/C assignments are deterministically randomized per state. Paired request interventions keep that state byte-identical. A fourth grade-0 passage, D, is reserved for the state intervention. The input contains only query text, local passage labels, and passage text; qrel grades, original passage identifiers, and scenario names are excluded. There are 276 states in total. Scenario balance is constructed, not an estimate of deployment prevalence, and the two years share the same English corpus.

The initial protocol was fixed after an earlier ranking exploration and before the reported runs; Qwen2.5-1.5B was added subsequently. Low output-alphabet support in the initial results motivated the explicit-output replication. The two realizations are analyzed separately.

\subsection{Interventions}
\begin{table}[t]
\caption{What each paired comparison changes. State includes the query and displayed passages.}
\label{tab:design}
\small
\begin{tabular}{@{}p{.25\columnwidth}p{.29\columnwidth}p{.34\columnwidth}@{}}
\toprule
Contrast & Changed & Fixed \\
\midrule
Request set & One vs. three questions & State, target proposition \\
Joint request & One vs. three emitted answers & State, target proposition \\
State addition & Append grade-0 D & Target question, A/B/C passage text \\
Document isolation & Remove other passages & Target question, target passage text, label \\
Eligibility & A/B vs. A/B/C criteria & Full A/B/C state \\
Abstention & Add N=None criterion & State, passage labels \\
Question order & Reverse question list & State, questions' content \\
Question keys & Rename routing keys & State, propositions \\
\bottomrule
\end{tabular}
\end{table}
The native audit contrasts all versus single requested questions and expanded versus original states through each released engine (Table~\ref{tab:design}). The joint-all versus joint-single comparison runs only for the ordinary-LLM controls. Additional arms reverse questions, add auxiliary questions, rename keys, isolate individual documents, change output types, and vary selection criteria. Auxiliary arms match the question count.

The ordinary-LLM controls yield 43 scored distributions per state and model: 47,472 per realization, or 94,944 across the initial and explicit-output runs, excluding preliminary and repeated-input checks.

\subsection{Open decision models}
Our primary open-decision group consists of Laya typed-decisions (Laya-typed; ModernBERT-based, approximately 421M parameters), Kev-4B (Qwen3.5-4B-Base with a trained adapter and pointer head), and Bespoke-Nimble-9B (a decision-finetuned Qwen3.5-9B adapter) \citep{laya2026,kev2026,nimble2026}. Repository visibility guided the initial selection, which required published task-trained weights and usable local inference for retrieval questions.

We use fixed checkpoint revisions and the released inference engines. Laya uses its native Noul/Choice/Score interface and shipped temperatures. Kev uses its FP32, unmerged reference evaluation path and shipped temperature 2.40605. Nimble uses its released Boolean/enum schema, BF16 merged adapter, and temperature 1.0 without test-set calibration. Its ordinal Score is derived from an ordered four-level enum. All models receive the same canonical state JSON and semantically identical self-contained questions in their required native schemas.

The native extension was designed after inspection of the ordinary-LLM results. The expanded group adds general-purpose Laya (Laya-general), Kev-0.8B, Kev-9B, and Von 1.2 \citep{von2026}. Its protocol was fixed before expanded-checkpoint inference, but after inspection of initial Laya results and partial Nimble and Kev runs; these analyses are exploratory. The first three checkpoints extend the initially selected families, while Von adds an option-isolated implementation. Each Kev scale uses its published base checkpoint. Von retains its shipped input-conditioned calibration and zero-shot Noul prior. Its public probabilities, like Laya's, are rounded to four decimals.

All seven checkpoints use the same 276 states, yielding 31 analyzed distributions per state and 59,892 in total. The 48,300 native requests also return auxiliary-question outputs. All analyzed inputs retain the complete evidence and criteria without truncation.

Laya-general exceeds its 512-token training budget on 14 states under at least one intervention, although all of them fit the supported encoder context. A matched sensitivity analysis excludes the affected queries in full. Laya-typed's longest input is 646 tokens, within its 1024-token training budget. Both Laya checkpoints use the encoder-supported 8,192-token context and a 256-token question budget. This matches Laya-typed's shipped budget and raises Laya-general's shipped budget from 192; all runs use the 256-token setting, and both values are recorded.

Native inference uses PyTorch 2.8.0 and Transformers 5.17.0. We refer to the saved fixtures, distributions, and analysis code collectively as \emph{the artifact}.

\subsection{Ordinary-LLM controls}
The control group comprises locally available Qwen3.5-4B, Qwen3.5-9B, and Llama-3.1-8B-Instruct, with Qwen2.5-1.5B-Instruct as the exploratory historical checkpoint. We use BF16, batch size one, Transformers 5.5.0, SDPA, no prompt truncation, and Qwen's non-thinking template. Every output label is verified to be one token.

The explicit-output realization appends the allowed labels and prohibits keys, quotes, JSON, and explanatory prose. For Qwen3.5 only, these reruns also use validated FLA/causal-convolution kernels instead of the original PyTorch fallback. Llama and Qwen2.5 retain their original environment. A three-state identical-input check found two argmax changes among 129 distributions after the kernel change. Paired analyses stay within each realization; the Qwen3.5 comparison across realizations combines wording and kernel changes.

\subsection{Aggregation and uncertainty}
We average target-level outcomes within each query and scenario, then weight the three scenarios equally in query-macro summaries. We average class-specific accuracy within each query and true class, then across queries. All bootstrap intervals resample queries 5,000 times with seed 20260928, keeping years separate. For a group with no observed events, the bootstrap interval degenerates to $[0,0]$; in tables reporting event-rate intervals, we mark those entries with $\dagger$ and omit that interval. Under independent sampling of queries from a common eligible-query population, $1-0.05^{1/n}$ gives the exact one-sided 95\% binomial upper bound on the probability that a query has at least one such event when none is observed in $n$ queries: 7.4\%/5.5\% for the two years. This query-level bound concerns the indicated flip or action-change event, not the judgment-level event rate or a signed accuracy difference. Appendix~\ref{app:nativethresholds} reports all three fixed thresholds for the native models.

\section{Results}
Paired results for the two years are reported in 2019/2020 order.
\subsection{Request and evidence dependence}
\begin{figure*}[t]
\centering\includegraphics[width=.94\textwidth]{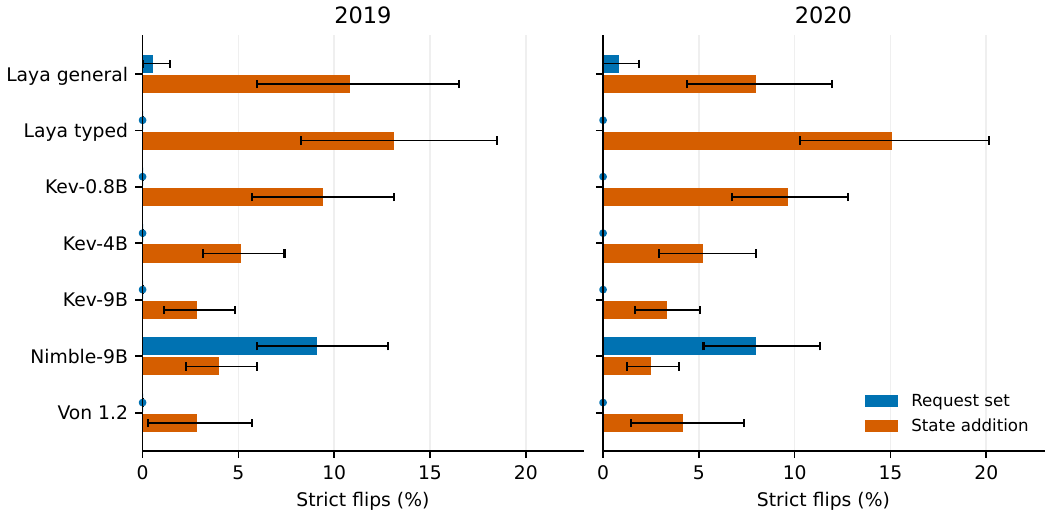}
\Description{Grouped bar chart of strict-flip percentages for the request-set and state-addition interventions across seven checkpoints and two years, with bootstrap whiskers; Nimble and Laya-general show nonzero request-set flips.}
\caption{Changing the requested questions versus adding a passage to the state. Bars show strict-flip percentages; whiskers are 95\% query-bootstrap intervals. Dots at zero mark zero observed flips, whose query-level upper bounds remain positive. Years and interventions are kept separate.}
\label{fig:dependencies}
\end{figure*}

\begin{table*}[t]
\caption{Native trained decision engines on the frozen retrieval states. Outcomes are percentages; brackets give 95\% query-bootstrap intervals where shown. Request and state columns measure strict binary flips; accuracy is single-question (singleton) Noul accuracy; action change is the percentage of states whose selected passage or abstention status changes when questions are batched at threshold 0.5. $\dagger$: no observed events; the degenerate bootstrap interval is omitted. The one-sided 95\% upper bound for a query having any corresponding event is 7.4\%/5.5\% (39/53 queries), a different estimand from the table's event rates.}
\label{tab:native}\centering\small
\begin{tabular}{llrrrr}\toprule
Decision model & Year & Request flips & State flips & Singleton accuracy & Action change \\\midrule
\tableinput native.tex
\bottomrule\end{tabular}
\end{table*}
Laya-typed, all Kev scales, and Von have no observed strict request-set flips, whereas Nimble-9B flips 9.1\%/8.0\% of target judgments (Table~\ref{tab:native}, Figure~\ref{fig:dependencies}). The corresponding mean absolute probability movement is zero at Von's four-decimal output resolution, below $10^{-6}$ for all Kev scales, and 0.102/0.095 for Nimble; Laya's intermediate movement is analyzed with the precision diagnostics below. In the released implementations, Laya encodes each question separately and Kev isolates question branches. Nimble scores each field in a separate pass but serializes the full request schema into every prompt.

\paragraph{Nimble field masking.}
We remove the other fields' definitions from each target's serialized schema while keeping the three-question request, question identifiers, state, and engine settings fixed. An unmasked rerun first reproduces every frozen probability. Every masked prompt is token-identical to the corresponding singleton prompt, and all 828 masked judgments reproduce the singleton probabilities exactly. All 70 observed request-set flips disappear. This intervention identifies the visible schema text as the source of request dependence on these states.

Two content-free auxiliary questions also flip 4.6\%/5.5\% of judgments (95\% query-bootstrap intervals 2.6--6.8 and 3.6--7.5). Thus extra fields can change a judgment even when they contain no competing-passage information. This auxiliary comparison is not length-matched.

\paragraph{Nimble flip directions.}
Batching moves 29 judgments from No to Yes and three from Yes to No in 2019; the corresponding counts are 32 and six in 2020. The net directional excess is 7.4 percentage points (95\% query-bootstrap interval 4.3--11.4) and 5.5 points (2.5--9.0). These flips comprise 16 corrections and 16 regressions in 2019, and 25 corrections and 13 regressions in 2020. The net task-quality changes from batching are 0.0 accuracy points (95\% query-bootstrap interval $-2.6$ to 2.8) and 0.0 correct-action points ($-6.0$ to 6.0) in 2019, versus $+2.5$ (0.2--4.6) and $+8.2$ (3.8--12.6) in 2020. In 2019, the 25 changed actions include ten corrections, ten regressions, and five changes with unchanged correctness (Appendix~\ref{app:nativethresholds}). The zero net change therefore reflects offsetting corrections and regressions.

Of Nimble's 25 changed actions in 2019 and 30 in 2020, 22 and 24 coincide with a strict flip in the same state. Abstention-to-selection changes (17 and 18) outnumber selection swaps and new abstentions.

Appending judged grade-0 passage D flips 13.1\%/15.1\% of Laya-typed decisions and 5.1\%/5.2\% of Kev-4B decisions. Nimble's rates are 4.0\%/2.5\%. For every checkpoint and year, removing the other passages changes more judgments than adding one: document isolation flips 27.4\%/25.4\% of Laya-typed, 12.5\%/11.5\% of Kev-4B, 22.8\%/19.9\% of Nimble, and 16.8\%/18.0\% of Von judgments.

\subsection{Precision and option positions}
Laya-typed's BF16 request contrast has mean absolute probability shifts of 0.00134/0.00151 and changes 1.7\%/1.9\% of software actions: two of 117 and three of 159 states. Both action-change intervals include zero (Table~\ref{tab:native}). All five changes swap accepted passages without changing abstention status or coinciding with a strict binary flip.

BF16 autocast is the pinned release's default on CUDA compute capability 8 or above, including our H800; CPU defaults to FP32. We use the CUDA default for the primary results and FP32 for the precision diagnostic.

\begin{figure*}[t]
\centering\includegraphics[width=.94\textwidth]{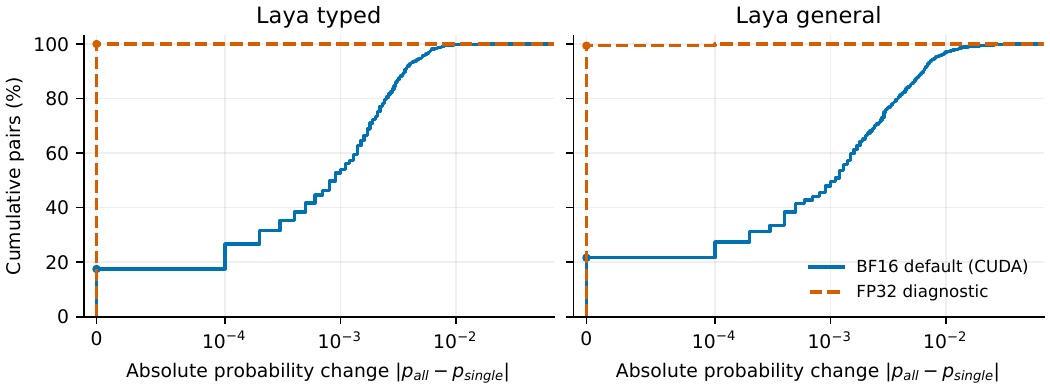}
\Description{Distributions of absolute request-induced probability movement over 828 paired judgments per Laya checkpoint, under the BF16 CUDA default and the FP32 diagnostic; nearly all FP32 mass sits exactly at zero.}
\caption{Empirical distributions of request-induced absolute probability movement over 828 pairs per Laya checkpoint (both years). BF16 is the native CUDA default; FP32 is a separate diagnostic run. The horizontal axis is linear near zero and logarithmic above $10^{-4}$. Dots show the mass at exactly zero: 17.5\%/21.6\% in BF16 and 100\%/99.4\% in FP32 for typed/general, where Laya-general's FP32 maximum movement is 0.0001. Probabilities are native four-decimal outputs.}
\label{fig:precision}
\end{figure*}

Figure~\ref{fig:precision} shows all 828 paired judgments per checkpoint. In FP32, Laya-typed has identical singleton/all probabilities at native four-decimal resolution. Laya-general has 0.6\%/0.8\% strict flips in BF16 (mean absolute shifts 0.00212/0.00237), versus none in FP32, where the maximum probability difference is 0.0001. The FP32 contrast supports a batch-shape/precision explanation for the small default-path movement. Laya-general's action changes are 0.0\% in 2019 and 2.5\% in 2020 (0.6--5.0), the only Laya action interval excluding zero (Table~\ref{tab:native}).

\paragraph{Option order and output types.}
\begin{table*}[t]
\caption{Native interface sensitivity (strict flips, \%). Question order/keys keep state fixed; binary Choice and Score are compared with Noul. The last two columns test conditional A/B preference under criteria reversal and A/B-to-A/B/C expansion, respectively, keeping the full A/B/C state fixed. The first four outcome columns are percentages of judgments; the last two are percentages of states. Full query intervals and tie counts are in the artifact.}
\label{tab:nativecontracts}\centering\small
\begin{tabular}{llrrrrrr}\toprule
Model & Year & Q order & Q keys & Binary Choice & Score & Criteria rev. & Eligibility \\\midrule
\tableinput native_contracts.tex
\bottomrule\end{tabular}
\end{table*}
Reversing Choice criteria changes approximately 5\% of Kev-4B's conditional A/B preferences (95\% query-bootstrap intervals 0.9--10.3 and 1.9--8.8), despite its zero observed request-set flips (Tables~\ref{tab:native} and~\ref{tab:nativecontracts}). Nimble's criteria-reversal rates are 14.5\%/15.1\%. Von's independent-option implementation has zero observed criteria-reversal flips and zero probability movement at reporting resolution.

Noul, binary Choice, and ordinal Score use different criteria encodings, priors, and calibration; Score also changes the semantic rubric. Von's zero-shot Noul applies a context-free polarity correction absent from its ordinary Choice path. The type-substitution comparisons measure the combined effect of these encoding and calibration choices.

\paragraph{Eligibility and separator position.}
Eligibility expansion flips conditional A/B preferences in 9.4\%/8.2\% of states for Laya-general, 4.3\%/5.7\% for Laya-typed, and 6.0\%/3.1\% for Nimble, with all six query intervals excluding zero, and with zero observed flips for all Kev scales (Table~\ref{tab:nativecontracts}). Zero strict flips can coexist with probability changes: Kev-4B's mean absolute conditional A/B shift is 0.00226/0.00211, and Von's is 0.0194 in 2019 (Appendix~\ref{app:nativeprob}). When C becomes eligible, Von has no observed conditional A/B preference flips in 2019 and two among 159 states in 2020 (query interval 0.0--3.1). Conditioning on A/B removes the mechanical softmax-denominator change. In the released code, the independent-option mask includes the terminal separator in the shared prefix, while the separator's rotary position depends on the total packed option length.

We test the separator's role on the two affected states plus the first eight states of the frozen 2019 enumeration as controls. A forward hook records native logits; a separate diagnostic fixes only the terminal separator's position relative to the prefix. The maximum A/B option-logit movement under eligibility expansion falls from 0.592 to $1.14\times10^{-5}$, and both preference flips disappear. On these selected states, the intervention identifies the shared separator's position as a source of eligibility dependence.

\subsection{Decision accuracy}
\begin{table*}[t]
\caption{Task quality with full shared evidence: native singleton Noul above the rule, ordinary-LLM controls (explicit-output realization, independent-compiler arm) below it. Accuracy and correct actions are percentages; Brier is unscaled. Actions use threshold 0.5. Overall gives the correct-action rate across all states; Zero/One/Two give rates for states with that many judged answer-bearing passages. All-No obtains 66.7\% accuracy; always abstaining obtains 33.3\% correct actions. Calibration differs: Kev uses shipped temperatures, while Nimble uses unfitted $T=1$.}
\label{tab:nativequality}\centering\small
\begin{tabular}{llrrrrrr}\toprule
Model & Year & Accuracy & Brier & Overall & Zero & One & Two \\\midrule
\tableinput native_quality.tex
\midrule
\tableinput native_quality_controls.tex
\bottomrule\end{tabular}
\end{table*}
Laya-typed's singleton accuracy is 66.1\%/64.8\%, close to the 66.7\% all-No baseline, while its action accuracy exceeds that of always abstaining (Table~\ref{tab:nativequality}). Its positive-class recall is 51.3\%/45.9\% and its balanced accuracy is 62.4\%/60.1\%; All-No has zero positive recall and 50\% balanced accuracy (Appendix~\ref{app:nativeclass}). Kev-4B reaches 75.2\%/80.1\% binary accuracy and 63.2\%/74.8\% correct actions. Nimble reaches 74.1\%/77.1\% binary accuracy and 58.1\%/62.3\% correct actions under the same threshold. It correctly abstains on approximately 85\% of zero-answer states, with positive-class recall of 37.6\%/41.5\%. Predicting the constructed positive prevalence, $p=1/3$, for every judgment gives a Brier score of $2/9\approx0.222$; Brier measures squared probability error and does not isolate calibration.

All three Kev scales have zero observed request-set flips. Their singleton accuracies are 51.0\%/53.2\% at 0.8B, 75.2\%/80.1\% at 4B, and 78.1\%/85.3\% at 9B. Von also has no observed request-set flips, with 37.0\%/40.9\% accuracy under its released Noul calibration. Kev-0.8B, Von, and the Llama control fall well below the 66.7\% all-No baseline (Table~\ref{tab:nativequality}). Laya-typed has higher correct-action rates than Laya-general in both years; its binary accuracy is lower in 2019 and higher in 2020.

Appendix~\ref{app:controls} compares the ordinary-LLM workflows. Appendices~\ref{app:nativeclass} and~\ref{app:nativethresholds} report native class-specific quality and all three thresholds. For Nimble in 2020, requesting all questions changes correct-action rates by $+8.2$, $+7.5$, and $+3.1$ points at thresholds 0.5, 0.7, and 0.9; the last interval includes zero. These fixed operating points retain each implementation's score scale and calibration. Eight model/year/threshold configurations abstain on every state; Appendix~\ref{app:nativethresholds} reports coverage and marks these configurations explicitly. Rounding public probabilities changes one Laya-typed reconstructed argmax among 1,104 Choice calls; retaining the native Choice winner avoids this discrepancy, which does not affect the Noul-based main tables.

\subsection{Sensitivity analyses}
\paragraph{Context-matched exclusion.}
The context-matched sensitivity analysis excludes all states from nine queries, retaining 37/46 queries and 249 states for every model. Request-set flip rates remain zero for all Kev scales, Laya-typed, and Von; they are 0.6\%/0.7\% for Laya-general and 9.3\%/7.7\% for Nimble.

\paragraph{Von polarity correction.}
We rerun all 828 Von singleton judgments, first reproducing every frozen native probability and then disabling only the additive context-free Noul correction. Accuracy rises from 37.0\%/40.9\% to 39.9\%/44.0\%; the paired gains, computed before rounding, are 2.8 points (95\% query-bootstrap interval 0.6--5.7) and 3.1 points (1.0--5.9). Yes rates fall from 95.2\%/89.5\% to 92.3\%/85.1\%. Removing the correction reduces excessive acceptance but leaves most errors intact. In this post-hoc diagnostic, the temperature function stays fixed, although its input-conditioned value can respond to the changed logits. Main tables retain native outputs.

\paragraph{SciFact request probe.}
The SciFact probe uses 20 scientific claims from BEIR \citep{wadden2020scifact,thakur2021beir}, selected by numeric query ID before inference. Each state contains one qrel-positive article and two TF-IDF-selected distractors, with complete titles and abstracts. The question asks whether an article supplies evidence relevant to assessing the claim, including support or refutation. Kev-4B and Nimble were selected because their TREC request contrasts had zero and nonzero observed flips, respectively.

On 60 singleton/all pairs per model, Kev-4B has zero observed strict flips, with a zero-event query upper bound of 13.9\%. Nimble has three flips across three claims (5.0\%, 95\% query-bootstrap interval 0--10.0): one No-to-Yes and two Yes-to-No; the interval includes zero. This exploratory probe reports request behavior only, because distractors outside the positive qrels are not verified negatives. The 240 additional distributions remain separate from the TREC totals.

\subsection{Ordinary-LLM controls}
With explicit output instructions, ordinary-LLM all-question label support exceeds 99.8\%. Request-set flips remain at 9.4--18.8\% and action changes at 21.4--49.1\%. Appendix~\ref{app:controls} reports support, correctness, joint-generation, and action results alongside the initial-output runs (Appendix~\ref{app:initial}) and numerical-kernel differences.

\section{Discussion}
The mechanism diagnostics distinguish separate computation from input isolation. Nimble evaluates fields in separate forward passes, yet each pass sees the full schema; masking that schema removes the observed request dependence. Von isolates options through its attention mask, but a shared separator's position still depends on the eligible set. Independence therefore depends on both the visible content and its positional encoding in these implementations.

Probability stability, binary stability, and action stability are distinct outcomes. Laya-typed's five action swaps occur above the binary threshold, whereas Nimble's request flips include both corrections and regressions. Action-change rates describe how often the decision rule changes its output; paired correctness differences describe the net task effect under the fixed rule and constructed states.

Reusing a judgment when output fields are added requires invariance to the request set (Table~\ref{tab:design}). That property does not establish invariance to evidence or eligibility changes, which alter different inputs. For retrieval applications, the relevant assessment combines these dependency checks with the selection and abstention rule and its task accuracy.

\section{Limitations}
The main study covers seven trained checkpoints from four decision implementations, plus four ordinary-LLM controls, on one English corpus, two test years, and constructed states with three initial passages. The hosted Jev service and official-adapter end-to-end behavior are untested. The SciFact extension covers request sensitivity for two models on 20 claims. Query eligibility, source-order selection, and the constructed scenario mixture limit transfer to deployment prevalence, risk, and calibration. Grade-0 labels may be imperfect, and new evidence can legitimately change a judgment, so state flips are not automatically errors. Exact deduplication leaves possible semantic duplicates; no new human adjudication was commissioned. Upstream training-data overlap, including SciFact, remains unverified.

Repository visibility guided selection, so the sample is not representative of open decision-model implementations. Differences in training data, base models, prompts, precision, context budgets, and calibration confound architecture and decision-training effects in cross-model comparisons, including the Kev size comparison. The ordinary-LLM compiler controls input visibility; Jev's training, calibration, execution efficiency, and native confidence measures remain outside that control.

The intervals quantify query-sampling variation, exclude training-run and model-family variation, and use no multiplicity adjustment. The additional checkpoints and the FP32, separator-position, field-masking, polarity-prior, and SciFact analyses are exploratory. Field masking identifies a channel on the existing fixture, and the separator diagnostic uses selected counterexamples; neither estimates the population accuracy of a retrained or repaired model.

Jev's interface does not require interchangeable probabilities across types or option sets. Token-label support is measured only for the ordinary-LLM controls, and their constrained probabilities differ from native confidence measures. Repeated-input checks establish reproducibility within the recorded configuration; invariance across hardware and libraries is untested. Shared GPU workloads and differing execution paths preclude a throughput comparison. The artifact is not yet publicly available; redistribution of source data is subject to the applicable licenses.

\section{Conclusion}
The audit identifies request dependence through Nimble's visible schema, small Laya probability shifts associated with numerical precision, and a Von eligibility dependency through separator positions on two affected states. Their consequences differ: Laya-typed's observed action swaps occur without binary flips, while Nimble's flips occur in both directions, with a positive net gain in correct-action rate in 2020 at thresholds 0.5 and 0.7. Across the constructed retrieval states, checkpoints with zero observed request flips differ substantially in positive-class recall, negative-class accuracy, and selection or abstention quality.

\bibliographystyle{ACM-Reference-Format}
\bibliography{../references,references}

\onecolumn
\appendix
\section*{Appendices}
\input{native_diagnostics.tex}
\input{controls.tex}
\subsection{Initial Output Instructions}\label{app:initial}
\begin{appendixtable}
\caption{Initial-output realization. Strict flip rates (\%) with 95\% query-bootstrap intervals. Interpret these alongside the alphabet-support diagnostic in Table~\ref{tab:support}.}
\centering\small\begin{tabular}{llrrr}\toprule
Model & Year & Request flips & Joint flips & State flips \\\midrule
\tableinput original.tex
\bottomrule\end{tabular}
\end{appendixtable}
The initial realization and explicit-output replication are analyzed separately.

\subsection{Request Order, Keys, and Output Types}\label{app:sensitivity}
\begin{appendixtable}
\caption{Explicit-output diagnostics: strict binary flips (\%, 95\% query-bootstrap intervals). Type comparisons use the independent compiler as the Noul baseline and include instruction/label changes.}
\centering\small\begin{tabular}{llrrrr}\toprule
Model & Year & Question order & Question keys & Noul vs. binary Choice & Noul vs. Score \\\midrule
\tableinput sensitivity.tex
\bottomrule\end{tabular}
\end{appendixtable}
The question-order arm reverses the visible question list; renaming changes the list's routing identifiers and the target reference. Both keep passage state fixed. The visible-schema proxy exposes routing keys, whereas native Jev keeps them outside the model input.

\subsection{Actions by Answer Availability}\label{app:scenario}
\begin{appendixtable}
\caption{Explicit-output correct actions (\%) by number of answer-bearing passages. Binary and ordinal workflows use threshold 0.5; Score uses mass on grades 2/3. Choice+None uses its categorical argmax. Correct action means abstention for zero answers and selection of any positive otherwise.}
\centering\small\begin{tabular}{lllrrrrr}\toprule
Model & Year & Answers & Compiler & Visible all & Joint all & Score & Choice+None \\\midrule
\tableinput scenarios.tex
\bottomrule\end{tabular}
\end{appendixtable}
Aggregate results weight the three constructed scenarios equally.
\end{document}

%% file: native_diagnostics.tex
\section{Additional Native-Model Analyses}
\subsection{Native Class-Specific Quality}\label{app:nativeclass}
Table~\ref{tab:nativeclass} reports the singleton Noul judgments used in Table~\ref{tab:nativequality}, with Yes defined by $p>0.5$. Positive-class recall (TPR) is the fraction of grade-2/3 targets accepted; negative-class accuracy (TNR, or specificity) is the fraction of grade-0 targets rejected. Each query contributes three positive and six negative judgments across the constructed states. Within-class accuracy is averaged per query, then across queries. Balanced accuracy is $(\mathrm{TPR}+\mathrm{TNR})/2$. All-No has TPR 0\%, TNR 100\%, and balanced accuracy 50\%.

\begin{appendixtable}
\caption{Native singleton judgments at threshold 0.5 (\%). TPR and TNR include 95\% query-bootstrap intervals; balanced accuracy (BA) is a point estimate. The constructed positive prevalence is one third.}
\label{tab:nativeclass}\centering\small
\begin{tabular}{llrrr}\toprule
Model & Year & Positive recall (TPR) & Negative accuracy (TNR) & BA \\\midrule
\tableinput native_class_quality.tex
\bottomrule\end{tabular}
\end{appendixtable}

Nimble's high negative-class accuracy and lower positive recall explain its frequent correct abstention and missed answers. Von exhibits the opposite pattern. Laya-typed's positive recall distinguishes its decisions from All-No despite their similar overall accuracy in this imbalanced construction.

\subsection{Probability Changes}\label{app:nativeprob}
Table~\ref{tab:nativeprob} reports probability movement for the request, evidence, and eligibility contrasts. Eligibility uses the conditional A/B probability; the other contrasts use the target's Yes probability. Zero strict flips indicate no observed pairs on opposite sides of 0.5; pairs that touch the threshold do not count as strict flips. All 117 Von states in 2019 change conditional A/B probability under eligibility expansion, with no strict preference flip. Laya-typed changes 683 of 828 probabilities in the request-set comparison and swaps five selected passages without a strict binary flip.

\begin{appendixtable}
\caption{Mean absolute probability change (unscaled, up to three significant digits). Changes are averaged within each query, then across queries. Request and evidence compare target Yes probabilities; eligibility compares conditional A/B probabilities. Laya and Von use their native four-decimal probability outputs.}
\label{tab:nativeprob}\centering\small
\begin{tabular}{llrrr}\toprule
Model & Year & Request & Evidence & Eligibility \\\midrule
\tableinput native_probability.tex
\bottomrule\end{tabular}
\end{appendixtable}

\clearpage
\subsection{Native Threshold Sensitivity}\label{app:nativethresholds}
Tables~\ref{tab:nativethresholds2019} and~\ref{tab:nativethresholds2020} use thresholds 0.5, 0.7, and 0.9 fixed before inference, with released calibration. The rule selects the highest-scoring passage above threshold or abstains. Requests use identical evidence; zero-, one-, and two-answer states receive equal weight.

Coverage is the passage-selection rate. In both years, Laya-typed at 0.7/0.9, and Laya-general and Von at 0.9, abstain on every state: eight configurations. Their 33.3\% correct-action rate comes from zero-answer states; every available answer is missed.

Paired $\Delta$ correct is the all-question correct-action rate minus the singleton rate, computed before rounding. Nimble's 25 action changes in 2019 at 0.5 include ten corrections, ten regressions, and five changes with unchanged correctness, yielding zero net change.

\begin{appendixtable}
\caption{Native threshold sensitivity on TREC DL2019 (39 queries, 117 states). Correct S/A denote singleton/all-question correct-action rates (\%); Coverage S/A denote passage-selection rates (\%); Action change gives the percentage of states whose selected passage or abstention status changes; $\Delta$ correct gives the paired difference (percentage points). Brackets are 95\% query-bootstrap intervals. $\dagger$: no observed action changes, hence zero observed accuracy difference; degenerate bootstrap intervals are omitted. The one-sided 95\% upper bound for a query having any action change is 7.4\%, not a bound on the signed accuracy difference. $\ddagger$: both requests abstain on every state.}
\label{tab:nativethresholds2019}\centering\small
\renewcommand{\arraystretch}{0.9}
\begin{tabular}{lrrrrrrr}\toprule
Model & Threshold & Correct S & Correct A & Coverage S & Coverage A & Action change & $\Delta$ correct \\\midrule
\tableinput native_thresholds_2019.tex
\bottomrule\end{tabular}
\end{appendixtable}

\begin{appendixtable}
\caption{Native threshold sensitivity on TREC DL2020 (53 queries, 159 states). Columns and intervals follow Table~\ref{tab:nativethresholds2019}. $\dagger$: no observed action changes and zero observed accuracy difference. The one-sided 95\% upper bound for a query having any action change is 5.5\%. $\ddagger$: both requests abstain on every state; Coverage S/A are singleton/all-question passage-selection rates.}
\label{tab:nativethresholds2020}\centering\small
\renewcommand{\arraystretch}{0.9}
\begin{tabular}{lrrrrrrr}\toprule
Model & Threshold & Correct S & Correct A & Coverage S & Coverage A & Action change & $\Delta$ correct \\\midrule
\tableinput native_thresholds_2020.tex
\bottomrule\end{tabular}
\end{appendixtable}

%% file: controls.tex
\section{Ordinary-LLM Controls}\label{app:controls}
\paragraph{Comparison with Kev-9B}
The Qwen3.5-9B independent-compiler control obtains 70.9\%/75.5\% correct actions, compared with Kev-9B's 65.0\%/76.1\%. These workflows differ in trained checkpoints, prompts, precision, and kernels, which confound the effect of decision training.
\subsection{Output support}
\begin{appendixtable}
\caption{Initial versus explicit-output realization. Alphabet mass is the full-vocabulary probability of allowed labels in the all-question probe. Request flips strictly cross 0.5. Values are percentages; brackets are 95\% query-bootstrap intervals. *Exploratory historical checkpoint. Qwen3.5 also changes kernels across realizations.}
\label{tab:support}\centering\small
\begin{tabular}{llrrrr}\toprule
Model & Year & Initial mass & Explicit mass & Initial request flips & Explicit request flips \\\midrule
\tableinput support.tex
\bottomrule\end{tabular}
\end{appendixtable}
Llama's initial all-question alphabet mass is 0.55\%/0.61\%, with strict flip rates of about 23\% in both years (Table~\ref{tab:support}). Its initial single-question mass is also low (8.50\%/9.39\%), whereas the joint output has more than 99.8\% mass on its permitted labels. Low label support confounds format-following effects with question sensitivity.

Explicit output instructions raise Llama's all-question support above 99.8\% and reduce its request flips to 9.4\%/11.5\%. All four models exceed 99.8\% all-question alphabet mass in this realization, and every model/year group retains request-set flips. The replication is a sensitivity analysis prompted by the initial label-support diagnosis.

\subsection{Request changes and accuracy}
\begin{appendixtable}
\caption{Explicit-output replication: strict binary flip rates and change in binary accuracy for the request-set contrast. Flips are percentages; accuracy changes are percentage points. Brackets are 95\% query-bootstrap intervals. Joint scores condition on generated prefixes; state scores use the independent compiler.}
\label{tab:results}\centering\small
\begin{tabular}{llrrrr}\toprule
Model & Year & Request flips & Joint flips & State flips & Request $\Delta$ accuracy \\\midrule
\tableinput results.tex
\bottomrule\end{tabular}
\end{appendixtable}
Under explicit output instructions, request flips range from 9.4\% to 18.8\% (Table~\ref{tab:results}). Qwen3.5-4B exhibits 17.1\%/12.8\% flips, and Qwen3.5-9B 16.0\%/15.9\%.

When other questions become visible, Qwen3.5-4B gains 6.6/5.0 binary-accuracy points and Llama gains 10.5/11.7 points. Qwen2.5-1.5B loses 4.3/7.3 points, although the 2019 interval includes zero. Qwen3.5-9B has changes of mixed sign, with both intervals including zero.

The independent compiler omits other questions from each target's input and is request-invariant by construction. Adding a grade-0 passage changes that input and induces 0.9--7.4\% flips in the explicit realization.

\subsection{Joint outputs}
Joint-all versus joint-single flips range from 17.0\% to 53.7\% in the explicit replication. Llama has the largest rates, 53.3\%/53.7\%. The joint prompt changes the requested output, and later labels also condition on previous decisions.

In the initial realization, Llama's first-output flip rates are 8.5\%/11.3\%, compared with 70.1\%/70.4\% for later outputs. Only later outputs condition on earlier answers, but target identity also varies with output position, so these rates do not isolate a prefix effect. Reversing joint order likewise changes both request serialization and preceding answers.

\subsection{Selection and abstention}
\begin{appendixtable}
\caption{Explicit-output replication, threshold 0.5. Left: selection-or-abstention action changes under request and joint interventions (\%, 95\% query-bootstrap intervals). Right: correct abstention on zero-answer states (\%).}
\label{tab:workflow}\centering\small
\begin{tabular}{llrrrrr}\toprule
Model & Year & Request change & Joint change & Compiler & Visible & Choice+None \\\midrule
\tableinput workflow.tex
\bottomrule\end{tabular}
\end{appendixtable}
Request changes alter 21.4--49.1\% of selection-or-abstention actions (Table~\ref{tab:workflow}). A flip for an unselected passage can leave the selected action unchanged.

A selector restricted to A/B/C must choose a passage in a zero-answer state; adding N makes abstention available. In the explicit replication, Llama's compiler never abstains correctly on zero-answer states despite its high alphabet support. Qwen2.5's compiler correctly abstains on about 79\% of those states, while its all-question visible probe never does. For Qwen3.5-9B, Choice+None correctly abstains on 48.7\%/67.9\% of zero-answer states. Appendix~\ref{app:scenario} compares all five workflows on zero-, one-, and two-answer states.

\subsection{Output types and request format}
Appendix~\ref{app:sensitivity} reports question-order, key-renaming, binary-Choice, and ordinal-Score comparisons. Key sensitivity arises in the visible-schema proxy; hiding keys removes that input channel.